\documentclass[aps,prl,amsmath,ams,groupedaddress,twocolumn,showpacs]{revtex4}
\usepackage{graphicx}
\usepackage{dsfont}
\begin{document}

\title{Quantum chaos in the spectrum of operators used in Shor's algorithm}

\author{Krishnendu Maity}
\author{Arul Lakshminarayan}
\email[]{arul@physics.iitm.ac.in}
\affiliation{Department of Physics\\ Indian Institute of Technology Madras\\
Chennai, 600036, India.}

\preprint{IITM/PH/TH/2006/3}

\begin{abstract}
We provide compelling evidence for the presence of quantum chaos in
the unitary part of the operator usually employed in Shor's factoring algorithm.
In particular we analyze the spectrum of this part after proper desymmetrization
and show that the fluctuations of the eigenangles as well
as the distribution of the eigenvector components follow the 
CUE ensemble of random matrices, of relevance to quantized chaotic
systems that violate time-reversal symmetry. However, as the algorithm 
tracks the evolution of a single state, it is possible to employ
other operators, in particular it is possible that the generic quantum chaos found above becomes of a nongeneric kind such as is found in the quantum cat maps,
and in toy models of the quantum bakers map.

\end{abstract}
\maketitle


\newcommand{\newc}{\newcommand}
\newc{\beq}{\begin{equation}}
\newc{\eeq}{\end{equation}}
\newc{\kt}{\rangle}
\newc{\br}{\langle}
\newc{\beqa}{\begin{eqnarray}}
\newc{\eeqa}{\end{eqnarray}}
\newc{\pr}{\prime}
\newc{\longra}{\longrightarrow}
\newc{\ot}{\otimes}
\newc{\rarrow}{\rightarrow}
\newc{\h}{\hat}
\newc{\bom}{\boldmath}
\newc{\btd}{\bigtriangledown}
\newc{\al}{\alpha}
\newc{\be}{\beta}
\newc{\ld}{\lambda}
\newc{\sg}{\sigma}
\newc{\p}{\psi}
\newc{\eps}{\epsilon}
\newc{\om}{\omega}
\newc{\mb}{\mbox}
\newc{\tm}{\times}
\newc{\hu}{\hat{u}}
\newc{\hv}{\hat{v}}
\newc{\hk}{\hat{K}}
\newc{\ra}{\rightarrow}
\newc{\non}{\nonumber}
\newc{\ul}{\underline}
\newc{\hs}{\hspace}
\newc{\longla}{\longleftarrow}
\newc{\ts}{\textstyle}
\newc{\f}{\frac}
\newc{\df}{\dfrac}
\newc{\ovl}{\overline}
\newc{\bc}{\begin{center}}
\newc{\ec}{\end{center}}
\newc{\dg}{\dagger}
\newc{\prh}{\mbox{PR}_H}
\newc{\prq}{\mbox{PR}_q}

The signatures of classical chaos in the quantum domain has been of
continuing interest for many years now and impacts various areas
of physics \cite{BerryQC,LesH,Haake}. The recent developments in quantum information theory 
and quantum computation has also prompted studies that delve on the
effects of chaos on quantum computers \cite{Shep00} and on entanglement \cite{Arul01}, a key 
resource in such processes. There have also been studies that seek
efficient implementation of quantum chaotic models on quantum 
computers \cite{Georg01}, as well as, to the best of our knowledge, one study that seeks to see if there is 
intrinsic chaos in some quantum algorithms \cite{Braun}. Such algorithms are
typically unitary evolutions, generated ultimately
by Hamiltonian evolutions, followed by measurements. The previous 
study \cite{Braun} focused on the quantum Fourier transform and Grover's search algorithm,
and several tests of quantum chaos were used. The evidence for quantum 
chaos was not unequivocal due to extreme degeneracies and other unusual
features. Besides, the quantum Fourier transform viewed as Weyl quantization,
quantizes a ninety degree rotation of phase space and should therefore
not be expected to have properties typical of quantum chaos. For any
value of dimensionality of the transform, its fourth power is unity.
 
On the other hand that Shor's factoring algorithm \cite{Shor} is a candidate for
quantum chaos has been indicated by earlier works of one of the
authors \cite{aruljphys}. This is due the fact that the order finding algorithm, 
at the heart of Shor's algorithm, has a key component, the modular 
exponentiation operator, which is 
essentially a shift permutation operator $S$. This shift permutation
operator has been shown to be metrically close to the quantum
baker's map \cite{BalVor}, quantization of a paradigm of classical chaos, namely
the double-sided left shift \cite{LL}. Operators closely allied with the shift operator  may also be thought of as quantizing a multivalued \cite{Nonnen} or a
  random map \cite{Scott}.
 Viewed as a Weyl quantization, its action on phase space coherent states 
produces stretching and folding \cite{aruljphys}, its overall periodicity makes it
akin to the quantum cat map \cite{BerryCat}. The quantum cat map quantizes another
classically fully chaotic system, the cat map \cite{ArnAvez}, however its quantum 
propagator is exactly periodic, with a periodicity that plays the role
of the order in Shor's algorithm. We have also previously shown 
how to construct the quantum baker's map using the shift operator 
and suitable projectors \cite{aruljphys}.

In this Paper we examine Shor's algorithm as a whole and show that 
its unitary part has properties that one would normally ascribe
to systems that are classically chaotic and for which time-reversal
symmetry is broken. The order finding part of the algorithm \cite{Shor} is 
quantum mechanical, and involves two registers containing $n_1$ and $n_2$
qubits respectively. We call the corresponding Hilbert spaces ${\cal H}^1$
and ${\cal H}^2$. The standard product basis on the space ${\cal H}^1\otimes
{\cal H}^2$ is denoted 
as $|j\kt |k \kt $, $0 \le j \le 2^{n_1}-1$ and $0 \le k \le 2^{n_2}-1$.
 Shor's algorithm proceeds by using the following 
operator:
\beq
U=  (F^{-1} \otimes \mbox{Id}) \, U_x\, (H \otimes \mbox{Id})
\eeq
Here $F^{-1}$ is the inverse discrete Fourier transform and $H$ is the
Hadamard matrix which act only on the first register, while $U_x$ is the entangling part defined by its 
action on a basis vector $|j\kt |k \kt$ as 
\beq
U_x|j \kt | k \kt = |j \kt |x^j k \, \mbox{mod}\, N \kt \equiv 
|j \kt \, S^j |k \kt, \;\; 0\le k \le N-1.
\eeq
If $k \ge N$ then $U_x |j\kt |k \kt = |j \kt |k \kt$. 
This defines the shift operator $S$ as $S|k\kt=|x\,k \,\mbox{mod}\, N
\kt$ for $0\le  k \le N-1$ and $S|k \kt =|k \kt$ otherwise. 
Here $N$ is the integer we wish to factor and $x$ is an integer that is co-prime to it.

For our study below we will take $x=2$ and $N$ to
be an odd integer so that we are guaranteed that an integer $r$ exists
such that $2^r = 1 \, \mbox{mod} \, N$, where $r$ is the order we are
seeking. Thus $U$ acts non-trivially in an $2^{n_1} \, N$ dimensional
subspace of the full Hilbert space ${\cal H}^1 \otimes {\cal H}^2$.
Now Shor's algorithm proceeds by taking a particular initial state 
$|0\kt |1\kt$, acting on this with $U$ and measuring the first register,
followed by classical steps intended to find the order $r$, from which
using standard number theory it maybe possible to find a factor of $N$ 
if it exists. We will analyze the {\em entire} spectrum of $U$ considered as
an operator of dimension $2^{n_{1}}\, N$. 

We note that as the order finding algorithm needs to consider only action on the initial state $|0\kt \otimes |1\kt$, $U$ is not the only operator that achieves the
necessary result. For instance the first operation of a Hadamard gate on the qubits of the first register maybe replaced by a Fourier transform, as acting on $|0\kt$ this also produces an equal superposition of all standard basis states: $H|0\kt=F|0\kt$. In this case the overall unitary part of the algorithm would be 
\beq
\tilde{U}=  (F^{-1} \otimes \mbox{Id}) \, U_x\, (F \otimes \mbox{id})
\eeq
The eigenvalues of $\tilde{U}$ are thus the same as that of $U_x$. The central operation is
the modular exponentiation and the quantum chaos in this can be made 
"generic" or not depending on the choice of unitary operators, such as $U$ or
$\tilde{U}$ above \cite{referee}. The experimental realizations of the Shor algorithm and order finding algorithms \cite{chuang} that have been carried out so far use the Hadamard gates on the first register, and the operator $U$ is of relevance herein. We will return to consider $\tilde{U}$ later, but for now consider the standard operator $U$.

We first notice that 
\beq
\left [ U, \mbox{Id}\otimes S \right] =0.
\eeq
We can label the eigenstates of $U$ with eigenvalues of $S$,
which are like good quantum numbers. The spectrum of $S$ is thus
of interest. As $S^r=\mbox{Id}_N$, we have 
\beq
S|s_j \kt = e^{i \theta_j} |s_j \kt, \, 0\le j \le N-1
\eeq
where $\theta_j$, the eigenangle,
is of the form $2 \pi k/r$ and $0 \le k \le r-1$.
The eigenstates of $U$ can be chosen to be $|\phi_l \kt |s_j\kt$,
unentangled states of the two registers. We show this 
as follows. Let $H|\phi_l \kt = \sum_{m} a_m |m\kt$ and 
$|s_j\kt = \sum_{k} b_k |k\kt$. Then 
\beqa
U_2 (H \otimes \mbox{Id}) |\phi_l \kt |s_j \kt=
U_2 \sum_{m,k} a_m \, b_k |m \kt |k \kt = \nonumber \\
\sum_{m,k} a_m \, b_k |m \kt S^m |k \kt=
\sum_m a_m e^{i m \theta_j} |m \kt |s_j \kt= \nonumber \\
\sum_m e^{i m \theta_j}|m \kt \br m |H |\phi_l \kt |s_j\kt 
= (\Lambda_j H  \otimes \mbox{Id})|\phi_l \kt |s_j \kt
\eeqa
where $\Lambda_j = \sum_m e^{i m\theta_j} |m \kt \br m |$ is 
a diagonal operator on the first register whose entries are
powers of the eigenvalues of $S$. Hence  
$
U|\phi_l \kt |s_j \kt = 
 (F^{-1}\Lambda_jH|\phi_l \kt ) |s_j \kt. 
$
Therefore $|\phi_l\kt |s_j \kt $ will be an eigenstate of $U$ with
eigenvalue $\lambda_{lj}$ if
\beq
F^{-1} \Lambda_j H |\phi_l \kt = \lambda_{jl} |\phi_l\kt,
 \; \; 0\le l \le 2^{n_1}-1.
\eeq
Thus we have split or block-diagonalized the full $2^{n_1}\, N$ dimensional
matrix diagonalization problem to that for $N$ matrices of dimensions
$2^{n_1}$ each. There is also a dependency of the eigenstates $|\phi_l\kt$
on the eigenangle $\theta_j$, but we suppress this. 

The operators $F^{-1} \Lambda_j H $, for $0\le j \le N-1$ are unitary
operators whose eigenvalues are that of the unitary part of Shor's
algorithm. When $\theta_j=0$, the relevant operator is simply
$F^{-1} H$. This ``Fourier transform of the Hadamard transform" was studied
recently as a model of eigenstates of quantum chaos \cite{NCNSD}. It was demonstrated
that columns of this matrix could be multifractals in the $N\rarrow \infty$
limit with peaks connected to the periodic and homoclinic orbits 
of the doubling map $x \mapsto 2 x \, \mbox{mod}\,1 $. These are of
relevance to the spectrum of the quantum baker's map.
It is thus of interest that a generalized construction arises in the
 spectral problem of Shor's algorithm. 

On using the matrix elements of $F^{-1}$ and the Hadamard matrix
it is possible to write the matrix elements $(F^{-1} \Lambda_j H)_{kl}=$\\
\beq
\f{1}{2^{n_2}}\prod_{m=0}^{n_1-1} \left( 1+ (-1)^{b_m} \,
e^{-2 \pi i k 2^{m-n_1}} e^{i \theta_j 2^m}   \right)
 \eeq
 where $l=\sum_{m=0}^{n_1-1} b_m 2^m$ is the binary representation 
 of $l$. When $\theta_j=0$ and $l=2^{n_1}-1$ corresponding the case
 $b_m=1$ for all $m$, this is the Fourier transform of the 
 Thue-Morse sequence \cite{Allpaper}, well-known to be a multifractal in the
 large $n_1$ limit \cite{Luck}. Thus the matrix elements of $F^{-1} \Lambda_j H$
 while having a simple form that is efficient to compute are in fact 
 quite complex objects. We now demonstrate that their spectrum has
  characteristics of that of a random matrix.
 
We illustrate this with a case: $n_1=10, N=29$. We diagonalize 
$F^{-1} \Lambda_j H$ for five different values of $\theta_j$,
namely $-20 \pi/28$, $0$, $4 \pi/28$, $6 \pi /28$ and 
$14 \pi /28$, choosing these to be a mixture of generic and special
eigenangles of $S$. The eigenvectors of $S$ can also be written for example as:
\beq
\label{vecS}
|s_j\kt = \dfrac{1}{\sqrt{r}}\sum_{n=0}^{r-1}\exp\left(\dfrac{-2 \pi i j n }
{r}\right)\, |2^n \, 
\mbox{mod}\, N \kt,
\eeq
where $0\le j \le r-1$ are eigenvectors corresponding to eigenvalues
$e^{2 \pi i j/r}$. In general this is not the complete set, but others
can be found based on subgroups generated by other ``seeds", where 
the seed is the integer $i_0$ and the group it generates is the set
of integers $2^k i_0 \, \mbox{mod} \, N$ for various $k$. For instance
in the case $N=29$, the above set generates $r=28$ eigenstates of $S$ 
with the seed $1$. The remaining state is a stand-alone one with the
seed $0$ and is the state $|0\kt$ itself, with an eigenvalue $1$. Thus
apart from the double degeneracy of this eigenvalue the other $27$ 
eigenvalues are non-degenerate. However this depends on the order $r$,
for example if $N=31$, $r=5$ the spectrum of $S$ is highly degenerate.
In these cases there are other symmetries like bit-flip that arise
 \cite{meenarul}, but we will not elaborate on these as they are inessential to the central
purpose of this Letter.
 
It is however pertinent to point out that eigenvectors
such as in Eq.~(\ref{vecS}) are completely delocalized, in fact have
modulus unity for almost all components, and the phase would seem
random. Thus these are simple examples of states that are ergodic in accordance with  
Shnirelman's theorem \cite{Schnil} about a measure of states that tend to be ergodic
in the classical limit for quantized ergodic systems. The classical limit
in this case would be over integers $N$ that are such that their order (with 
respect to 2) is $N-1$.  

\begin{figure}
\includegraphics[width=2.5in,angle=-90]{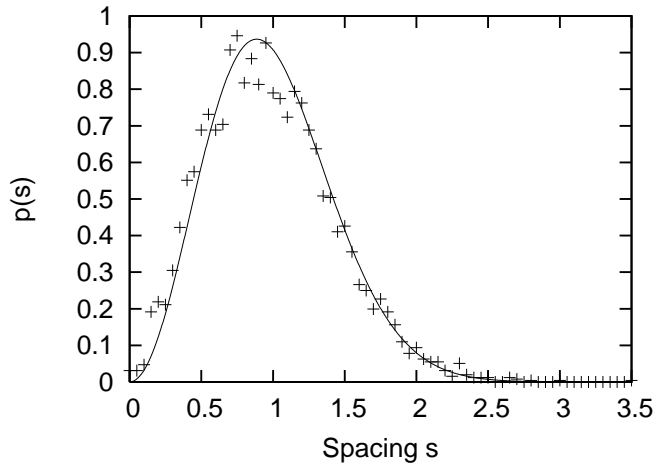}
\caption{The nearest neighbor spacing distribution of eigenangles 
from an ensemble consisting of 5115 level spacings for the case when the first
 register has $10$ qubits and the number to be factored is $29$. The smooth curve
 shows the CUE distribution of random matrix theory.
}
\label{nns}
\end{figure}

Returning to the central issue, we find the nearest neighbor
spacings (nns) of eigenangles for the five chosen cases thus making
an ensemble with statistical significance. The nns is calculated for
the normalized spacings $\Delta \alpha_{jl} 2^{n_1}/2\pi$ such that the
mean spacing is unity, where $\lambda_{jl}=e^{i \alpha_{jl}}$ and 
$\Delta \alpha_{jl}$ refers to spacings between nearest neighbours.
In Fig.~(\ref{nns})
we show how the nns is distributed along with the curve expected for
the Circular Unitary Ensemble (CUE), which consists of the unitary group
U(n) of $n \times n$ unitary matrices endowed with its Haar measure \cite{Mehta}.
The good agreement with the CUE distribution \cite{Mehta} (which coincides with the Wigner
surmise for the Gaussian ensembles for large dimensionality) 
\beq
p(s)= \df{32s^2}{\pi^2} e^{-4 s^2/\pi}
\eeq
indicates the applicability of random matrix theory to the spectral fluctuations
of the unitary part of Shor's algorithm. It is generally accepted that while there 
are exceptions, random matrix fluctuations are quantum signatures of classical chaos \cite{Haake}. 
In this case the classical limit may be considered to be the large $N$ (or $n_1$) limit,
which is in fact the regime where Shor's factoring algorithm will ever be usefully
implemented. 

\begin{figure}
\includegraphics[width=2.5in,angle=-90]{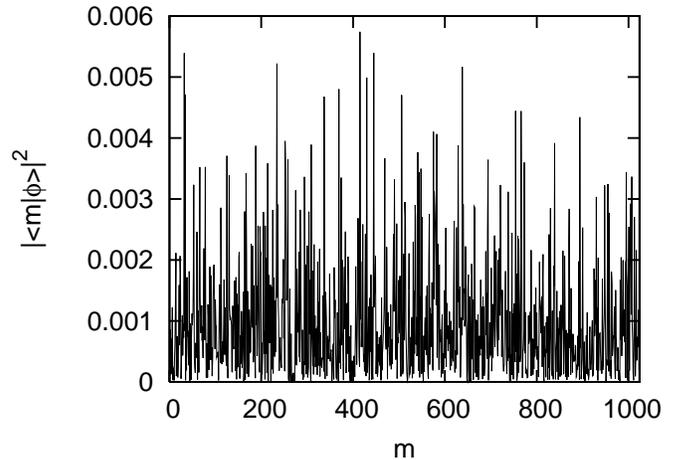}
\caption{The intensities of a typical eigenstate of the operator $F^{-1}\Lambda_j H$
for the same case as in Fig.~(\ref{nns}). The complete eigenstate of the unitary part
of Shor's algorithm being a tensor product of such states with eigenstates of the
shift permutation operator.}
\label{eigfns}
\end{figure}

\begin{figure}
\includegraphics[width=2.5in,angle=-90]{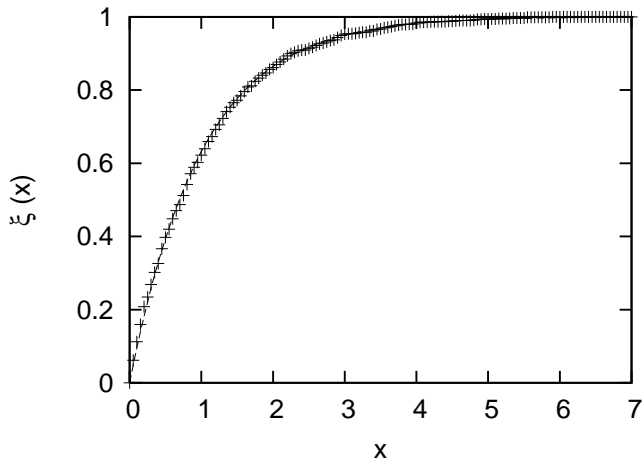}
\caption{The cumulative distribution $\xi(x)$ of the intensities of the eigenstate shown
in Fig.~(\ref{eigfns}), these being normalized so that the mean is unity. Shown as a 
smooth curve is the random matrix theory expectation $1-e^{-x}$.}
\label{efstat}
\end{figure}

The eigenfunctions are also of interest and in Fig.~(\ref{eigfns}) is shown a typical 
eigenstate of $F^{-1} \Lambda_j H$ in the standard basis for the first of the five values of
$\theta_j$ stated above. Almost all of the eigenfunctions have this very random appearance
and an analysis of the distribution of its normalized intensities $x= 2^{n_1} \, |\br m |\phi\kt |^2$
fits with that expected from random matrix theory. These normalized intensities with unit mean 
are distributed in an exponential manner $e^{-x}$. In Fig.~(\ref{efstat}) we show the 
cumulative distribution, $\xi(x)$, of $x$ and compare it to that expected from random matrix theory,
namely $1-e^{-x}$ \cite{Haake}, and again find a good fit. Data not shown here confirm both the 
nns and eigenvector component statistics for a variety of other parameter values
and states, the results shown being typical. Of course the complete eigenstate of the 
unitary part of Shor's algorithm is a tensor product of such eigenstates of random appearance
 with eigenstates of the form in Eq.~(\ref{vecS}), which have phase complexity but in 
 modulus are almost equidistributed.

Thus there is compelling evidence that the operator $U$ used in standard implementations of Shor's algorithm has quantum chaos in it, of the type expected of systems that do not have time-reversal symmetry. The genesis of this is from two sources, one is the modular exponentiation operator, which as we have noted earlier is closely allied to models of quantum chaos such as the quantum baker's map \cite{aruljphys}. The other is from a combination of the Fourier and Hadamard transforms. Thus the spectral properties
of $F^{-1}H$ by itself maybe interesting. It maybe also noted that this is the operator
relevant to the subspace $\theta_j=0$, which also includes the subspace left out due to the
modular exponentiation part acting as identity on it (of dimension $(2^{n_2}-N)\, 2^{n_1})$.
This last combination maybe made irrelevant to Shor's algorithm by making use
of $\tilde{U}$ instead of $U$. The operator $\tilde{U}$ is exactly periodic as both
$F$ and $U_x$ are. Its spectrum is highly degenerate and the same as the shift operator.
The eigenvalues are thus equally spaced on the unit circle, reminiscent of the quantum cat maps. The use of $H$ instead of $F$ ($U$ instead of $\tilde{U}$) seems to lift this nongeneric spectrum into a more generic one. There could be other operators that also
accomplish order finding with different initial states, but the core of the algorithm, the
modular exponentiation, will introduce quantum chaos into the system.

There could be implications of the RMT fluctuations found on the practical functioning
of the algorithm. In particular quantum chaotic systems have been found to have hypersensitivity to perturbations of the Hamiltonian or noise \cite{Schack,Peres}. It is possible that in some way the state used in Shor's algorithm as the initial state is ``protected" from this, but it remains to  be seen whether this is indeed the case. For this an analysis that concentrates on the time evolution rather than stationary states will be of relevance.
In particular it is of interest to investigate whether $U$ and $\tilde{U}$ are qualitatively
different in their response to perturbations, and more generally whether use of the Fourier
to produce equal superpositions out of $|0\kt$ instead of the Hadamard gate is more robust. Work is on in these directions \cite{meenakshi}.

\acknowledgements{ AL wishes to thank Arvind for discussions.}

\end{document}